\documentclass[12pt]{article}
\usepackage{amssymb,epsf,color}
\usepackage{amsmath}
\usepackage{bbm}
\def\lsim{\mathrel{\rlap {\raise.5ex\hbox{$ < $}}
{\lower.5ex\hbox{$\sim$}}}}
\def\gsim{\mathrel{\rlap {\raise.5ex\hbox{$ > $}}
{\lower.5ex\hbox{$\sim$}}}} 
\def\sqr#1#2{{\vcenter{\vbox{\hrule height.#2pt

        \hbox{\vrule width.#2pt height#1pt \kern#1pt

           \vrule width.#2pt}

        \hrule height.#2pt}}}}

\def\lsim{{\displaystyle
{{\raise-8pt\hbox{$ <$}}
\atop{\raise5pt\hbox{$\sim$}}}}}
\def\gsim{{\displaystyle
{{\raise-8pt\hbox{$ >$}}
\atop{\raise5pt\hbox{$\sim$}}}}}
\def\slsim{{\displaystyle
{{\raise-8pt\hbox{$\scriptstyle <$}}
\atop{\raise5pt\hbox{$\scriptstyle \sim$}}}}}
\def\sgsim{{\displaystyle
{{\raise-8pt\hbox{$\scriptstyle  >$}}

\atop{\raise5pt\hbox{$\scriptstyle \sim$}}}}}

\newskip\humongous \humongous=0pt plus 1000pt minus 1000pt

\newcommand{\sumpf}[0]{\sum_{(H^{\rm f},G^{\rm f})}\! \! \! \!
{\raise
4pt
\hbox{$'$}}\,}

\newcommand{\sump}[0]{\sum_{(H,G)}\! \! {\raise 4pt \hbox{$'$}}\,}

\def\bs{\begin{subequations}}
\def\es{\end{subequations}}

\catcode`\@=11
\newcount\hour
\newcount\minute
\newtoks\amorpm

\hour=\time\divide\hour by60
\minute=\time{\multiply\hour by60 \global\advance\minute by-\hour}
\edef\standardtime{{\ifnum\hour<12 \global\amorpm={am}%
        \else\global\amorpm={pm}\advance\hour by-12 \fi

        \ifnum\hour=0 \hour=12 \fi
        \number\hour:\ifnum\minute<10 0\fi\number\minute\the\amorpm}}
\edef\militarytime{\number\hour:\ifnum\minute<10 0\fi\number\minute}
\def\draftlabel#1{{\@bsphack\if@filesw {\let\thepage\relax
   \xdef\@gtempa{\write\@auxout{\string
      \newlabel{#1}{{\@currentlabel}{\thepage}}}}}\@gtempa
   \if@nobreak \ifvmode\nobreak\fi\fi\fi\@esphack}
        \gdef\@eqnlabel{#1}}
\def\@eqnlabel{}
\def\@vacuum{}
\def\draftmarginnote#1{\marginpar{\raggedright\scriptsize\tt#1}}
\def\draft{\oddsidemargin -.2truein
        \def\@oddfoot{\sl preliminary draft \hfil
        \rm\thepage\hfil\sl\today\quad\militarytime}
        \let\@evenfoot\@oddfoot \overfullrule 3pt
        \let\label=\draftlabel
        \let\marginnote=\draftmarginnote
   \def\@eqnnum{(\theequation)\rlap{\kern\marginparsep\tt\@eqnlabel}%
\global\let\@eqnlabel\@vacuum}  }
\def\subequations{\refstepcounter{equation}%
  \edef\@savedequation{\the\c@equation}%
  \@stequation=\expandafter{\theequation}
  \edef\@savedtheequation{\the\@stequation}
  \edef\oldtheequation{\theequation}%
  \setcounter{equation}{0}%
  \def\theequation{\oldtheequation\alph{equation}}}

\def\endsubequations{\setcounter{equation}{\@savedequation}%
  \@stequation=\expandafter{\@savedtheequation}%
  \edef\theequation{\the\@stequation}\global\@ignoretrue
  \vspace*{-12pt} \\}

\def\bs{\begin{subequations}}
\def\es{\end{subequations}}
\relax
\def\thefootnote{\fnsymbol{footnote}}
\def\be{\begin{equation}}
\def\ee{\end{equation}}
\def\ba{\begin{eqnarray}}
\def\ea{\end{eqnarray}}

\def\ee{\end{equation}}
\def\bea{\begin{eqnarray}}
\def\eea{\end{eqnarray}}
\def\nn{\nonumber}

\newcommand{\uarrw}[0]{\mathrel{
{\raise.5ex\vbox{\hrule width 1cm}\hskip-6pt\rightarrow}}}
\def\thebibliography#1{%
\vskip 0.5cm \centerline{\bf References}
\list{%
[\arabic{enumi}]}{\settowidth\labelwidth{[#1]}
\leftmargin\labelwidth
\advance\leftmargin\labelsep
\usecounter{enumi}}
\def\newblock{\hskip .11em plus .33em minus .07em}
\sloppy\clubpenalty4000\widowpenalty4000
\sfcode`\.=1000\relax}

\renewcommand{\theequation}{\arabic{section}.\arabic{equation}}

\renewcommand{\section}{\setcounter{equation}{0}\@startsection%
{section}{1}{0mm}{-\baselineskip}{0.5\baselineskip}%
{\normalfont\normalsize\bfseries}}

\renewcommand{\subsection}{\@startsection%
{subsection}{2}{0mm}{-\baselineskip}{0.5\baselineskip}%
{\normalfont\normalsize\slshape}}

\renewcommand{\subsubsection}{\@startsection%
{subsubsection}{2}{0mm}{-\baselineskip}{0.5\baselineskip}%
{\normalfont\normalsize\slshape}}

\begin{document}
%
%
\renewcommand{\theequation}{\arabic{section}.\arabic{equation}}
\begin{titlepage}
\begin{flushright}
\end{flushright}
\begin{centering}
\vspace{1.0in}
\boldmath

{ \large \bf On the critical temperatures of superconductors:\\
a quantum gravity approach\\
\bf }

\unboldmath
\vspace{1.5 cm}

{\bf Andrea Gregori}$^{\dagger}$ \\
\medskip
\vspace{1.2cm}
{\bf Abstract} \\
\end{centering} 
\vspace{.2in}
We consider superconductivity in the light of 
the quantum gravity theoretical framework introduced in~\cite{assiom}.
In this framework, the degree
of quantum delocalization depends on the geometry of the
energy distribution along space.
This results in a dependence of the critical temperature
characterizing the transition
to the superconducting phase on the complexity of the structure
of a superconductor.
We consider concrete examples, ranging from low to high temperature 
superconductors, and discuss how the critical temperature can be predicted
once the quantum gravity effects are taken into account.

\vspace{6cm}

\hrule width 6.7cm
\noindent
$^{\dagger}$e-mail: agregori@libero.it

\end{titlepage}
\newpage
\setcounter{footnote}{0}
\renewcommand{\thefootnote}{\arabic{footnote}}

\tableofcontents

\vspace{1.5cm}

\noindent

\section{Introduction}
\label{intro}

Since its discovery in 1911 by Onnes 
superconductivity 
constituted one of the main domain of research of theoretical and applied 
physics. It cannot be explained within classical mechanics,
and it has proven to be a test for the predictive power of quantum 
mechanics. Its fundamental theoretical explanation was
given in 1957 by Bardeen, Cooper and Schrieffer \cite{BCS}, 
as due to the non-locality of
electron wavefunctions which, under certain conditions, 
form bosonic pairs (Cooper's pairs). 
At sufficiently low temperature, the latter 
collapse to a narrow energy width that, 
according to the Heisenberg Uncertainty Principle, implies a high
delocalization in space.   
In the last decades superconductivity received new 
attention after the discovery of higher and higher temperature 
superconductors, opening the possibility of utilization
of superconductors for practical purposes.

One aspect of high temperature superconductors that leaps to the eyes is the 
intriguing correlation that there seems to exist 
between critical temperature of transition to the superconducting phase, and 
the complexity of the lattice corresponding to a superconducting 
crystal \footnote{See for instance the list of the table 
of page~\pageref{tableSc}.}. 
The grounds of a possible relation between lattice complexity 
and critical temperature remain however obscure. 
In the BCS theory, 
the critical temperature is related to the specific energies of the
phonon excitations of the superconductor. In this way, it is possible to
justify critical temperatures of some degrees Kelvin. Since the highest phonon 
frequency does not increase with an increase of the lattice length/complexity,
this explanation seems to fail in the case of high temperatures.
All these analyses are based on a quantum mechanical approach that
does not consider gravity as a phenomenon whose quantum aspects should give a
significant contribution.
Indeed, it is common belief that gravity should
play a negligible role in atomic and molecular phenomena, especially when
one deals with objects at rest and not explicitly subjected to any kind 
of (unbalanced) gravitational force, either because the system is not 
accelerated, or, as in the case of a laboratory experiment on the earth, 
because the gravitational acceleration is negligible as compared to the 
strength of the other forces in the game. However, as a matter of fact
lattice complexity means somehow geometric complexity of the 
mass/energy distribution along space (in this case a lattice 
element), and, according to General Relativity, 
talking of geometry of the energy 
concentration means somehow talking of gravity.
Indeed, although irrelevant at the atomic scale
from a classical point of view, gravity can play a role when quantized. 
In fact, quantization of gravity implies much more than just quantizing
the graviton field: the latter is only the effect which is seen perturbatively.
However, as is known, quantum gravity is not perturbatively stable as a 
quantum field theory. In this work,
we consider it in the light of the
non-perturbative quantum gravity framework proposed in Ref.~\cite{assiom}.
Within this theoretical framework,
once considered non-perturbatively, gravity can be shown to
produce a deep modification of the whole quantum world, at any scale.   
In this work, we argue that quantum gravity plays a role also
in the correlation between lattice complexity and critical temperature
of superconductors.

\vspace{1cm}

In Ref.~\cite{assiom} I proposed a theoretical scenario in which the universe, 
namely, the whole detectable physical content of the universe, 
is, at any time,  the result of the superposition 
of all possible configurations, i.e. distributions,
assignments, of units of a degree of freedom, that we call "energy",
along a target space of any possible dimension, 
that we call "space of positions". In its basic formulation, these 
spaces are discrete, so that these assignments
are basically assignments of occupation numbers of cells, like binary strings 
of information, that tell whether a cell at a certain position in the target 
space bears a unit of energy or not. 
If we consider the set $\left\{   \Psi \right\}_E$ consisting of all such 
configurations (distributions) at finite amount of total energy $E$, we can 
introduce an ordering through these sets, given by the amount of energy.
As $\left\{   \Psi \right\}_{E^{\prime}}  \supseteq 
\left\{   \Psi \right\}_{E}$ if
$E^{\prime} \geq E$, the value of $E$ can be considered like a time 
coordinate, and the $E$-ordering
through the sets $\left\{   \Psi \right\}_E$ the time ordering through the 
history of the universe. 
All this sounds rather abstract, but, once these abstract
configurations and their combinatorics are translated into the ordinary words 
and language of physics,
they can be shown to produce the "right" universe. 
The fact that what we observe is a superposition of 
configurations can be expressed by introducing a "generating function" for the 
observables:
\be
{\cal Z}_E ~ = ~ \int {\cal D} \psi \, {\rm e}^{S} \, .
\label{Zsum}
\ee 
This is the sum over all the configurations $\psi$ 
at a given total energy $E$, weighted by their volume
of occupation in the phase space of all the configurations 
$W(\psi) = \exp S(\psi)$ (the entropy $S$ is
defined in the usual way, through $S \equiv \log W$). The sum~\ref{Zsum} is 
taken over
an infinite number of configurations, and, as such, in general diverges. 
The mean values of observables
are however defined in a non-singular way, as:
\be
\langle  {\cal O} \rangle ~ \equiv ~ \frac{1}{\cal Z} \int {\cal D} \psi \, 
{\rm e}^{S} {\cal O} \, .
\label{ZsumO}
\ee
In this framework, causality is substituted by an evolution which, instead of 
being deterministic,
is rather "determined": at any time, the average appearance of the 
universe is predominantly given by 
the configurations of highest entropy. This gives in the average a 
three-dimensional universe
with the geometry of a three-sphere whose radius, after appropriate conversion 
of units, is $R \sim E$.
One can show that the uncertainty introduced in neglecting
more "peripherical" configurations, namely, configurations of lower entropy,
precisely corresponds to the Heisenberg's 
position/momentum,
or, better, time/energy, uncertainty (see Ref.~\cite{assiom}). 
The Heisenberg 
uncertainty principle can be considered as the ground statement of 
quantum mechanics, which can be viewed as the theoretical 
implementation of the Heisenberg's uncertainty through a probabilistic
wave-like description of physical phenomena. The way this uncertainty
arises in this scenario leads to 
the interpretation of quantum mechanics as a practical way of 
parametrizing the undefinedness
of the observables beyond a certain degree of approximation. Indeed,
according to \ref{ZsumO}, the mean value of an observable
receives contribution from an infinite number of configurations.
In particular, in \ref{ZsumO} are contained also
an infinite number of configurations given by distributions of 
energy degrees of freedom along
any space dimensionality, and, in general, escaping any interpretation in 
terms of the usual concepts of (smooth) geometry. 
The bound set by the Uncertainty Principle is here not simply a bound on the 
possibility of measuring certain quantities, but the ``threshold'' beyond 
which these quantities can not even  be defined, because 
space, time, energy, momentum are only average, mean concepts that can be
introduced only at a relatively large scale, averaging over an infinite number
of configurations.

One can also show that in this scenario
the maximal speed at which information propagates is the speed of expansion of 
the universe itself:
$v_{\rm max} \propto dR / dE (=t) = {\rm constant}$. 
This constant can be identified with the 
speed of light, and
called $c$ (see Ref.~\cite{rel}). Moreover, it can also be shown that the 
geometry of the distribution 
of energy is also the geometry of the trajectories of propagation. 
This theoretical framework describes therefore a quantum-relativistic scenario 
which, under appropriate limits
to the continuum, can be well approximated by a string-theoretical 
description. The description in terms of string theory is however not a 
fundamental physical property in itself, but a representation useful for 
certain purposes, such as the determination of the physical content implicit 
in~\ref{Zsum} in terms of spectrum of elementary particles, fundamental 
interactions, and the corresponding masses and couplings \cite{spi}.
For the purposes of our present analysis, mapping to a string theoretical 
description is not necessary. What interests us here is that,
in this theoretical framework, the Heisenberg's uncertainty
is the better and better satisfied as an equality 
by more and more ``smooth'' physical systems.
The energy uncertainty is bounded from below by the uncertainty
of the most ``classical'' configuration, the one of a three-sphere of radius
$c \Delta t$, and energy content $\sim \left(M^2_{\rm Pl} c^4 / \hbar \right)
\Delta t$ (this is also the basic geometry of the universe itself 
\cite{assiom,spi}): in this theoretical scenario
we can say that the fact of being also gravity quantized
reflects in the fact that the degree of 
non-classicity of a physical system
turns out to depend on its``geometry'', intended in the general relativistic
sense of space distribution of energy.
More complex quantum systems show a higher degree of 
quantum delocalization.

Quite a few physical systems look almost like a three sphere
with almost the energy density of a black-hole. But, as long as we look
at sufficiently extended bodies
with a big mass, uncertainties in position and momentum, 
even if not precisely at the ``equality bound'', are nevertheless
negligible. Things become more critical as we go to the
atomic, and subatomic, scale. In that case, the difference
between our theoretical framework and the usual 
approach becomes relevant.
Superconductors are typical systems in which this phenomenon becomes critical 
and evident. These are materials which, although
in themselves can be of huge extension, we are going to probe in their
small scale properties. And, the more, by probes, the electrons, in a 
rather non-classical regime, such as the collective Cooper-pairs wave functions
close to their ground energy. That is, where 
energy-momentum/time-position uncertainties play a relevant role, and where
therefore quantum gravity effects show up more evidently.
Apparently, this contradicts the popular idea that
quantum gravity should become relevant only at the Planck scale.
However, as it as discussed in \cite{assiom} and \cite{spi}, 
quantum gravity effects don't show up only at the Planck scale, but also at a 
much lower scale. In particular, according to \cite{spi}, also the masses
of the elementary particles, and in particular the electron's mass and the 
electroweak scale, find their natural explanation not within
field theory, but as quantum gravity phenomena.

\section{Quantum gravity and superconductors}

The phenomenon of superconductivity is explained in its grounds as due
to the formation of pairs of electrons, that, behaving 
thereby ``collectively'' 
as bosons, can fulfill a narrow band of energy obeying to Bose-Einstein
statistics. In other words, there can be very many within such a narrow
band, so to produce a non negligible electric current \cite{BCS}.
In the BCS argument essential for the occurring of this process is the
existence of an attractive potential, attributed to the
phononic response of the atoms of metal under charge displacement due to a 
motion of the electrons. This produces an energy gap $\Delta$, 
and it has been shown that at the critical 
temperature most of the electrons pairs lie in an energy range of order
$\Delta$, at an energy which lies a gap $\Delta$ above the Fermi energy.
This at least in the most simple formulation of the theory. More complicated
structures of superconductors require modifications of this simple model,
and eventually also weakens the existence 
of an energy gap as an essential feature of superconductivity, because
there are conditions under which superconductivity exists
even without an energy gap. We will not consider here the details of
these model modifications and adjustments, which, as in any attempt
to describe real, complex physical systems, are somehow unavoidable.
For what interests our present discussion, it is important to consider that,
whatever the derivation and the approximation introduced to reproduce a
physical model can be, superconductivity remains related 
to the existence of a sufficient amount of electrons
possessing a sufficient degree of non-locality. We can summarize
this by introducing a ``critical length'' $\xi$, which, for reasons
that will become clear in the following, does not necessarily coincide
with the ``coherence path'' it is usually talked about in the literature about 
superconductivity. For the time being, let us just assume that, according
to a certain mechanism, which can reasonably be the one of phonon response
of the BCS approach, Cooper's pairs do form and collect to a characteristic 
length higher than $\xi$ when the temperature is sufficiently low.
It must be stressed that the distribution of electrons is not a mathematical
step function. Step functions are useful approximations introduced in
practical computations. In reality, it is a matter of statistics. Therefore,
one should never forget that, at any temperature, there will be a certain 
amount of pairs with typical length below $\xi$, 
and a certain amount above $\xi$. The relative amounts are a matter of
temperature. 
If we call $n$ the total number of electrons
and $n_S$ the number of electrons which are paired and with typical
length larger than $\xi$,
we can define the critical temperature $T_c$ independently
on the possible existence of an energy gap, just as the temperature at which
$n_S/n \geq \left( n_S /n \right)_0$, $\left( n_S /n \right)_0$ being 
a certain well defined ratio, which
does not need to be better specified. $\xi$ is therefore a mean quantity. In 
traditional quantum mechanics, where gravity is switched off, $\xi$
can only increase as a consequence of a higher localization in the space of
momenta: $\langle \xi \rangle \sim \hbar / \Delta p \sim \hbar v_F / \Delta E$.
In our quantum gravity scenario, $\langle \xi \rangle$ depends instead 
also on the complexity of the geometry of the system. 
We want to see how this comes about, 
and how, as a consequence, the critical temperature too 
will turn out to depend 
on the complexity of the geometry.

\vspace{1cm}

As discussed in Ref.~\cite{assiom}, 
the uncertainty of the energy of a system during a time $\Delta t$ is:
\be
\Delta E ~ \gtrsim ~ \frac{\Delta t}{\Delta S_0} \, ,
\ee 
where $\Delta S_0$ is the
maximal possible entropy of a region created, ``existing'', for a time 
$\Delta t$.
The ordinary formulation of the Heisenberg's uncertainty relation follows
from expressing $\Delta S_0$ as $(\Delta t)^2$, the entropy of a black-hole
of radius $\Delta t$ (in units for which we set $c$, the speed of light, 
to 1). 
This gives a {\it lower bound} on the energy uncertainty, and is almost 
satisfied as an equality only by configurations (i.e. physical systems 
corresponding to configurations)
with a geometry/energy content close to that of a black hole.
As we are going to discuss now,
for configurations more remote, more ``peripherical'' in the phase space, 
the uncertainty will be higher.

Let us consider
the sum \ref{Zsum}. It describes a universe ``on shell''; namely,
the universe ``as it is''. This means that there is no isolated system
(particle or complex system of any kind), not embedded in its environment. 
In particular, there is no isolated system
existing in a flat space. Not only the dominant configuration
always contains the ground curvature of the universe, but any 
configuration $\psi$ involved in the sum \ref{Zsum} is a distribution
of $E = {\cal T}$ total energy degrees of freedom along a target space.
Any configuration describes therefore a ``whole universe''.
If we want to consider just a particular system, 
we must make an abstraction, and 1) look at just a subset of the
configurations contributing to \ref{Zsum}, 2)
for any configuration of this subset, we must restrict our
attention to a subregion of space.
There is here a subtlety, because, in general, $\psi$
does not describe a universe in three dimensions. As we said, this is true only
for the dominant configurations. On the other hand, if it is true that 
the contribution of non-three dimensional, less entropic configurations
is precisely what makes of the universe, and, in particular, of any subregion 
of it, a quantum system, 
it is also true that, in the concrete cases we want to consider here,
a full bunch of these configurations, and precisely the most entropic ones,
describe an energy distribution in a three dimensional space.
Otherwise, we would not be able to talk of superconductors in the terms
we are used to, namely, as well identified and (macroscopically) localized
materials in a three-dimensional space. 
The physical systems we consider are therefore 
``at the border'' between two descriptions: 
not anymore completely classical, but not even absolutely
remote in the phase space, in order to completely escape the ordinary 
parameters of our perception of a three-dimensional space-time, and 
therefore of operational definition through 
a set of measurement and detection rules and 
experiments \footnote{Indeed, as discussed in \cite{assiom}, in
this theoretical framework, precisely
quantization, i.e., the implementation of the Heisenberg uncertainty,
allows a description of a basically multi-dimensional world in
terms of three-dimensions, plus an uncertainty, an ``error'', under which
are collected the contributions of other dimensions.}.

Let us concentrate our attention on just a small part the universe, 
a piece of superconducting material and, possibly, its close environment,
with its atoms, electrons,
magnetic fields etc., namely, all what constitutes our ``experiment''. 
Let us call 
$E^{(sc)}$ the energy of this portion of the universe.
Of course, $E^{(sc)} < E$, the total energy of the universe (indeed, obviously
$E^{(sc)} \lll E$).
Let us consider the bunch of configurations of the universe that
contain our superconductor, $\left\{ \psi^{(sc)}  \right\}$.
Of course, for what we said \emph{all} the configurations contributing to
\ref{Zsum} do contain also the portion of universe in which our superconductor
is placed. However, what we want to do here is to select the subset of
configurations that contribute in a non-trivial way to form up the shape of the
superconductor, not just those that contribute, say, 
for the ground curvature of space.

When we measure the energy of our experiment,
the quantity that we detect is a mean value of energy,
$\langle E^{(sc)}  \rangle$, defined as
\footnote{All this can be put on a formal ground, by introducing 
an appropriate operator that, as is usual to do 
in the case of any generating functions, 
extracts from the logarithm of
\ref{Zsum} the energy of a space domain
around our experiment, 
but we don't want to bother here the reader with formalisms, 
rather to give the insight into the physical meaning of what we are doing.}:
\be
\langle E^{(sc)}  \rangle ~ = ~ \frac{1}{\cal Z} 
\int_{\psi \in \left\{ \psi^{(sc)}  \right\}} {\cal D} \psi
\; {\rm e}^{S} \, E^{(sc)} \, . 
\label{Emean}
\ee 
Let us consider how energy can be distributed
in the space, in order to form up our experiment of mean energy 
$\langle E^{(sc)}  \rangle$.
According to \ref{Zsum}, the more a configuration is remote in the phase 
space, the less it weights in the sum out of which we should compute the 
mean total energy of the experiment.
Since in a finite region of space we can arrange only a finite amount of energy
(we can put at most one unit of energy per each unit of space, where units of 
energy are measured in terms of Planck mass, units of space in terms of 
cells of Planck length size), to get a certain amount
of mean total energy $\langle E^{(sc)}  \rangle$ 
we must sum up over a larger and larger number of configurations.
The larger and larger,
the more and more remote the average configuration we want to describe.
Moreover, since in a finite region of space we can arrange only a finite 
number of different configurations of energy,
as we go further with the remoteness, to sum up to the same fixed 
amount of local energy $\langle E^{(sc)}  \rangle$ we must include 
configurations $\psi^{(sc)}$, in which $E^{(sc)}$ is
supported in larger and larger space regions. 
In terms of traditional quantum mechanics, this means that the wavefunctions
are more and more spread out in space.

As discussed in \cite{assiom} and \cite{spi}, configurations can be classified
according to the (finite) symmetry group of the 
distribution of energy degrees of freedom in the target space they correspond 
to. Their weight in \ref{Zsum} corresponds, by definition, to the number
of times they occur in the phase space, in turn given by 
the number of equivalent ways they can be formed.   
The ratio of the weights in the 
phase space of two configurations can be expressed as: 
\be
{W(\psi_i) \over W(\psi_j)} ~ = ~
{|| G_i  || \over || G_j ||} \, ,
\label{WWhh}
\ee
where $G_i$ and $G_j$ are the symmetry groups, and $|| G ||$ indicates the
volume of the group. This means that
the more symmetric a configuration is, the higher is its
weight in the phase space~\footnote{As discussed 
in \cite{assiom}, the basic definition of
space is discrete. Therefore, one
work always with finite groups, for which 
$G_i \neq G_j \Leftrightarrow || G_i || \neq || G_j ||$.}.
If a configuration $\psi_j$ corresponds to a more broken symmetry group
than a configuration $\psi_i$, it will be
more remote, more ``peripherical'' in the phase space.

Let us introduce the concept of mean weight of our experiment, and
of mean volume
of the symmetry, or volume of the symmetry group of the mean configuration, 
through:
\be
\langle W^{(sc)}   \rangle ~ = ~ {1 \over {\cal Z}} 
\int_{\psi \in \left\{ \psi^{(sc)}  \right\}} {\cal D} \psi
\, {\rm e}^{S} \, , 
\label{Wmean}
\ee 
and:
\be
{\langle || G^{(sc)}_i  ||  \rangle  \over
\langle || G^{(sc)}_j  ||  \rangle}
~ = ~ 
{\langle W^{(sc)}_i    \rangle  \over
\langle  W^{(sc)}_j    \rangle}
\, .
\label{Gmean}
\ee 
Accordingly, we define the mean configuration 
$\langle \psi^{(sc)} \rangle$ as the configuration 
for which:
\be
W \left( \langle \psi^{(sc)} \rangle  \right)  ~ \stackrel{\rm def}{=} ~
\langle W^{(sc)}  \rangle  
\, .
\label{psimean}
\ee
Let us suppose we change the symmetry of the configuration 
of our superconductor, 
$\langle \psi^{(sc)} \rangle \to \langle \psi^{(sc) \prime} \rangle$,
so that
$|| G^{(sc)}   || \to || G^{(sc) \prime} ||  
= {1 \over 2} || G^{(sc)} ||$. In order to build up the same
amount of energy $\langle E^{(sc)}  \rangle$ we must consider configurations
that distribute a higher amount of energy. 
How much more energy should we add, and how larger must be the space support?
Approximately we must consider twice as much energy, and,
in order to distribute it, at first sight we would say
we need twice the volume.
However, this second statement is not true.
The point is that the maximal \emph{mean} 
energy we can distribute along space, $\langle E \rangle$, goes linearly
with the radius, not with the volume.  
To keep fixed the ratio of volumes of
the symmetry groups, we must preserve 
the ratio of the entropies as compared with
the maximal entropy, which is the one of a black hole of radius
$\langle 2 E \rangle$.
Therefore, to maintain unchanged the value of $\langle E^{(sc)}  \rangle$ and 
$|| G^{(sc) \prime} || $, we need to consider distributions of twice as much
energy along a region with twice the radius, not the volume.
That is, energy goes \emph{linearly} with the space length.  
Therefore, the spreading in space of wavefunctions is
inversely proportional to the volume of the mean symmetry group:
\be
{\langle \Delta x \rangle \over \langle  \Delta x \rangle^{\prime}}
~ = ~ { \langle  || G^{(sc) \prime} || \rangle \over  
\langle  || G^{(sc)} || \rangle}
\, .
\label{dxG}
\ee
For any set of configurations with local symmetry group
$G_i$, we may think of $G_i$ as the little group of symmetry surviving
after quotienting a larger group $G$ through $h_i$. If
we have two sets of configurations, $\psi_i$ and $\psi_j $, obtained by
quotientation from the same initial group: $G_i = G/h_i$, $G_j = G / h_j$, 
we have:
\be
{W(\psi_i) \over W(\psi_j)} ~ = ~
{|| h_j  || \over || h_i ||} \, .
\ee
Passing from the generic $\psi_i$, $\psi_j$ to
$\psi^{(sc)}$, $\psi^{(sc) \prime}$, and introducing correspondingly
$h^{(sc)}$, $h^{(sc) \prime}$ instead of $h_i$, $h_j$, 
we can write \ref{dxG} as:
\be
{\langle \Delta x \rangle \over \langle  \Delta x \rangle^{\prime}}
~ = ~ { \langle  || h^{(sc)} ||\rangle \over  \langle || h^{(sc) \prime} || 
\rangle}
\, .
\label{dxh}
\ee
Since we are working at fixed mean energy $\langle E^{(sc)}  \rangle$,
we are allowed to consider that, in both primed and unprimed situations,
we have the same energy (or momentum) uncertainty.
Indeed, as discussed in \cite{assiom}, the energy uncertainty is given
by the contribution of the more peripherical configurations, with respect
to those over which one takes the average in order to obtain the mean energy.
Here, the energy uncertainty can be re-absorbed into a 
redefinition of mean volume of symmetry group, and therefore 
implicitly \emph{assumed} to be fixed, in order to derive the scaling of the 
space (and time) spread out. 
Under these conditions,
\ref{dxh} tells us that $\langle || h^{(sc)}  ||  \rangle$ can be viewed
as an \emph{effective} Planck constant:
\be
{ \langle \Delta x \rangle \times \langle \Delta p \rangle \over
\langle \Delta x \rangle^{\prime} \times \langle \Delta p \rangle  }
~ = ~ {\langle || h^{(sc)}  ||  \rangle \over 
\langle || h^{(sc) \prime}  ||  \rangle   } ~ \equiv ~
{h_{\rm eff}  \over h_{\rm eff}^{\prime} } \, .
\label{heff}
\ee
Up to an overall proportionality constant, that can be set to one, 
we can therefore write the quantum gravity version of the
Heisenberg Uncertainty as:
\be
\Delta x \Delta p ~ \geq ~ {1 \over 2} \hbar_{\rm eff} \, ,
\label{xpeff}
\ee
where $\hbar_{\rm eff} \equiv h_{\rm eff} / 2 \pi$ is related to
the symmetry of a configuration through \ref{heff}, \ref{dxh}, \ref{dxG}
and \ref{Gmean}.
Since increasing $|| h ||$ corresponds to increasing the complexity
of the configuration, 
things work as if the system would become less and less classical, 
more and more quantum mechanical, 
as the complexity of its structure increases.

We have identified
the critical temperature of superconductivity $T_c$ as 
the temperature at which a well defined portion of electronic-bosonic 
states are delocalized at least as much as a critical length $\xi$.
It is not necessary here to go into the details of the
actual computation of $T_c$ within a specific model.
It is enough to know that it is obtained by integrating over a statistical
distribution of states, and that 
the latter is expressed in terms of weights depending on $E / k T$.
Since everything depends on the ratio 
$E/T$, a rescaling of $E$ is compensated by a rescaling of the 
temperature $T$ while keeping fixed the ratio $E / T $.
In other words, $T$ can be viewed as the unit of measure of $E$.
To be concrete, let us consider once again 
our example of the two configurations characterized 
respectively by 
$|| G^{(sc)} ||$ and $|| G^{(sc) \prime} || = {1 \over 2} || G^{(sc)} ||$.
In the primed case, the same delocalization in space as in the unprimed
configuration corresponds to one-half
the unprimed energy. Since both energies are effectively ``measured'' 
in units of $T$, instead of talking of half energy,
we can speak of doubling the temperature. Coming back to the general case,
consider that
in \ref{xpeff} the effective Planck constant can be viewed
both as setting the scale of length as compared to energy/momentum, 
or equivalently as setting the scale of energy/momentum as compared
to space, and time.  
The relation \ref{heff} tells us therefore that, for more complex 
configurations,
the same amount of electrons with space delocalization $\xi$ will be obtained
at a higher critical temperature, according to:
\be
{T_c(i) \over T_c(j)} ~ = ~
{h_{\rm eff}(i) \over h_{\rm eff}(j) }
\, . 
\ee
In our theoretical framework, high critical temperatures 
show up as the consequence of the fact that,
as expressed in~\ref{dxh},
in superconductors with more 
complex geometrical structure, wavefunctions
have a larger quantum uncertainty. In particular, 
keeping fixed all other parameters, they have a larger $\langle \xi \rangle$.
Therefore,
the condition $n_S / n \geq (n_S / n)_0$ is satisfied
at higher temperature.

We stress that the considerations about the introduction of an effective,
geometry-dependent Planck constant concern the delocalization of wave
functions. Namely, the role the Planck constant plays in the Uncertainty
Relations, $\Delta x \Delta p \geq h / 4 \pi$, 
$\Delta t \Delta E \geq h / 4 \pi$, not the value of this constant as a 
conversion unit between energy and time, or space and momentum,
in contexts not related to the Uncertainty Relations 
\footnote{In other words, we could introduce a function of the geometry, 
which is set to
one for flat geometry (or, to better say, for the ground geometry of the 
universe, corresponding to a curvature $R$ of the order of the cosmological 
constant $\Lambda$ (see Refs.~\cite{spi} and~\cite{assiom}):
\be
\Delta x \Delta p \; \left[ \Delta t \Delta E \right]
~ \geq ~ {h \over 4 \pi} f(\langle R \rangle) \, , 
~~~~f(\langle R \rangle = \Lambda) \, = \, 1 \, .
\ee 
In this way, the Planck constant remains formally invariant, while the ratios
of above are expressed as ratios of different values of the function $f$.}. 
For instance, the
energy levels as computed through the Schr\"{o}dinger equation, or a set
of Schr\"{o}dinger equations,
out of a classical description of effective potentials, are computed
using the ground value of the Planck constant.
On the other hand, once the energy eigenvalues of a system are known, a 
geometry-dependent Planck constant must be used, in order to obtain
the effective spreading of wavefunctions in a geometrically complex 
quantum system.
To this regard, a consideration about the size of characteristic 
lengths which are 
introduced in the physics of superconductors, such as the coherence lengths
$\xi_0$, $\xi(T)$, and the London
penetration length $\lambda$, is in order. 
One could have the
impression that, as we are keeping fixed the critical delocalization
of wavefunctions at the transition to the superconducting phase,
the entire classification about what are type I and what type II
superconductors, discriminated by the ratio $\lambda / \xi_0$,
has to be reconsidered.
Indeed, this is not true, and all the classical results to this regard
go through 
unchanged in this scenario, because $\lambda$ contains in its definition
the Planck constant. In other words, both lengths $\lambda$ 
and $\xi_0$ scale in the same way, and,
as long as it is a matter of working with effective descriptions of 
superconductivity, such as for instance the Ginzburg-Landau effective theory, 
one can safely ignore rescalings, together with the grounds of 
a rescaling of the critical temperature.

\vspace{.6cm}

\section{Critical temperatures in various superconductors}

We consider now various examples of superconductors, in order
to see up to which extent
an approach based on our effective quantum gravity scenario can be applied.
The considerations of the previous section give us a clue on the role
played by the geometry of a superconductor in determining its critical 
temperature. However, the detection of a regime of superconductivity 
is in general
not a direct observation in itself: this regime is stated after observation
of several properties, such as for instance the magnetic properties.
Magnetic effects play a relevant role also in the generation of an effective
resistivity. Therefore, superconducting regime, and critical temperature
in particular, may be very sensitive to effects such as impurities, 
and in general doping effects aimed to pin magnetic vortices. Also external 
conditions such as pressure do play in general a significant role.
All these conditions affect of course also the ``geometry'' of the physical 
configuration under experiment from the quantum gravity point of view, but
quite often their most macroscopic effect reflects in the properties
of superconductivity not as a consequence of a modified geometry, but 
as a consequence of
a change in the dynamic magnetic properties or alike. Our investigation
is therefore affected by a large amount of imprecision, and must be taken
more as the indication of a tendency, than as a real precision test.

Low temperature superconductors are metals without a well defined
``structure''. As mentioned in the introduction, in this case
critical temperatures
are, in their order of magnitude, well predicted within the BCS theory. Our
concern will be with ``structured'' configurations, leading therefore
to higher critical temperatures. On the other hand,
according to our previous discussion, in
our approach we do not obtain an absolute
determination of critical temperatures, but only of their ratios, as 
a function of the ratios of geometries. For our analysis, 
we will therefore take
as a reference point the low BCS temperatures. 
As a starting point we 
take mercury, which has a critical temperature around 4,2 $^{\rm o}$K.
The reason why we consider this element instead of others is that
it allows a simpler derivation of the ratio of weights, in the sense 
of~\ref{Gmean}, to the next material we want to consider, NbTi.

\subsubsection{Hg $\to$ NbTi}

As a first test of the idea let us consider NbTi, the first step 
above Hg in the list of the table of page~\pageref{tableSc}. 
In first approximation, the structure
of NbTi should correspond to a $Z_2$ breaking of the symmetry of Hg. This would
be exactly true if Nb and Ti had the same mass and properties. 
Indeed, we can ideally consider the symmetry breaking as roughly
occurring through the pattern:  
\be
^{80}{\rm Hg} \, \stackrel{\sim Z_2}{\longrightarrow}
\, 2 \; ^{41}{\rm Nb} \, 
\stackrel{\mathbbm{1} \times Z_2}{\longrightarrow} \, ^{41}{\rm Nb} 
\, +
\, ^{22}{\rm Ti} \, +  \, 
^{22}{\rm Ti} \hspace{-.8cm}////
\, .
\ee 
This is somehow in between $Z_2$ and $Z_3$: it has less symmetry than a 
$Z_2$, but more than a $Z_3$, in that Nb looks like twice Ti, so that it
ideally comes from the recombination of a $Z_2$ symmetry subgroup out of a 
breaking into three Ti.
The critical temperature of NbTi should therefore lie somehow between
2 and 3 times the one of Hg: 
$T_c ({\rm NbTi}) \sim (8,4+ 12,6) / 2 \sim 10,5 ^{\rm o}$K. 
Indeed, the observed critical temperature lies around 10 $^{\rm o}$K.  
Of course, our evaluation has to be taken only as a rough, indicative estimate.

\
\\

In passing from Hg to NbTi we have introduced 
a ``weighted'' breaking of the Hg molecular symmetry. The weight is precisely
the mass of the atoms into which the initial homogeneous energy distribution
breaks. This is justified by the fact that, in
our theoretical framework, the configurations $\psi$ in \ref{Zsum} and 
\ref{ZsumO} are configurations
of energy distributions along space. The size of the
mass of a particle
depends on the weight the configuration (or the set of configurations) in which
this particle appears has in the phase space of all the configurations.  
In turn, the weight of a configuration depends on the symmetry of the 
\underline{energy} distribution. 
Approximately, the latter is ``measured'' by the space gradient of energy:
roughly speaking
the higher is the density of energy gradient, the less homogeneous
(= less symmetric) is the energy distribution.
This can be understood as follows: 
let us consider a configuration, i.e. a 
particular distribution of energy along space, in its fundamental definition,
as given in \cite{assiom}, namely, as a map from a discrete space to a 
discrete space. At any time we move one energy unit (unit energy cell in the 
language of \cite{assiom}) from a position
in the target space to a neighbouring one,
we modify one symmetry group factor. If we
move just one cell we increase (or decrease) the energy gradient
by two units, and break (restore) one ``elementary'' group factor.
If we move another unit, we increase (decrease) further
the energy gradient by two units, 
and act once again on another elementary group factor, and so on.
The amount of increase/decrease of the energy gradient is therefore
proportional to the factor of increase/reduction of the symmetry group
of the configuration.
These considerations are 
true for the configurations $\psi$ entering in \ref{Zsum} and
\ref{ZsumO}. However, owing to the properties of factorization of
the phase space, and assuming that such a factorization is a good approximation
when we want to ``isolate'' a local experiment such as those we are 
considering, we can transfer these
global considerations also to the local description of superconductors.
This implies neglecting the
``extremely peripherical'' configurations, anyway contributing for
a minor correction, negligible for our present purposes.
Therefore, instead of working with configurations as in \ref{Zsum} and 
\ref{ZsumO}, we work with ``averaged'' configurations as in \ref{psimean},
considering that everything outside the portion of universe
we are testing remains unchanged.
If we view configurations through an isomorphic representation
in terms of symmetry groups:
\be
\psi \, \leftrightarrow \, \prod_j G^{\psi}_j \, ,  
\ee
passing through the decomposition into an external and local part of the group:
\be
\prod_j G^{\psi}_j  \, = \, \left( \prod_j G^{\psi \, (ext)}_j \right)
\times
\prod_j G^{\psi \, (local)}_j \, ,
\ee
it becomes clear that each configuration $\psi_{\alpha}$
can be factorized as:
\be
\psi ~ = ~ \psi^{(ext)} \times \psi^{(local)} \, ,
\label{psiei}
\ee
where ``$local$'' and ``$ext$'' precisely mean respectively
the part of the configuration
(or the corresponding symmetry group) describing the experiment 
(superconductor and related environment), and the rest of the universe.  
We can translate these considerations
in terms of weights. Through the association:
\be
\langle \psi^{(sc)} \rangle 
\, \longleftrightarrow \, \langle W^{(sc)} \rangle \, = \,  
\int_{\psi \in \{ \psi^{(sc)} \}} {\cal D} \psi \, 
{\rm e}^S
\left( \prod_i G^{\psi \, (ext)}_i 
\prod_j G^{\psi \, (local)}_j \right) \, ,
\ee
we use the factorization \ref{psiei} to first integrate over the external 
part of every configuration. 
As long as the portion of universe represented by our 
experiment is very small as compared to the rest of the universe,
external and local part of configurations
can be approximately treated as independent. Under this approximation,
also the measure of integration can be factorized:
\be
{\cal D} \psi \, \longrightarrow \, {\cal D} \psi^{(ext)} \times 
{\cal D} \psi^{(local)} \, .
\ee 
We can therefore write:
\be 
\frac{1}{\cal Z} \int_{\psi \in \{ \psi^{(sc)} \}} {\cal D} \psi \,
{\rm e}^S
\left( \prod_i G^{\psi \, (ext)}_i 
\prod_j G^{\psi \, (local)}_j \right) \, \approx \,
\left\langle \prod G^{(ext)} \right\rangle \times 
\langle G^{(local)} \rangle \, ,
\ee
which allows to associate to $\langle \psi^{(sc)}\rangle $
a decomposition of weights:
\be
\langle \psi^{(sc)}\rangle 
\, \to \, 
\langle W^{(sc)}  \rangle
\, \approx \,
\langle W^{(ext)} \rangle \times \langle W^{(local)} \rangle
\, .  
\label{Wfact}
\ee
This decomposition allows us to reduce the analysis of
symmetries of configurations to just 
the crystal structure of our superconductors. 
The more, since superconductivity occurs as a property related to a 
characteristic length $\xi$, our considerations can be restricted to
a region of this extension. 
In general, it is enough to look at a scale of order of the lattice length:
the electron energy levels are given  
in terms of collective wave functions 
\footnote{For a review of these topics see for instance \cite{tinkham}.}, and
all quasi-particle energies are measured in terms of the
lattice length $a$, which sets 
therefore the effective length/energy scale.
In particular,
when the energy gradient between neighbouring lattice periods is sufficiently
``smooth'', it is possible to restrict the analysis to one lattice period.
This is the case of the majority of the examples we are going to consider.
With a certain degree of approximation, we can therefore write:
\be
\langle W^{(local)} \rangle ~ ~
\propto ~~  \approx \, \int_{a} | \nabla E_i |_{a} 
\, .
\ee
\ref{Wfact} allows us to write then:
\be
{\langle W^{(sc)} \rangle_j \over 
\langle W^{(sc)} \rangle_i} \, \approx \,
{\int_{a_i} | \nabla E_i |_{a_i} 
\over
\int_{a_j} | \nabla E_j |_{a_j} } 
\, \approx \,
{h_{\rm eff} (i) \over h_{\rm eff} (j)} \,
\approx \, 
{T_c (i) \over T_c(j)} 
\, .
\label{nablaE}
\ee
Since we are talking of elements basically at rest, we can 
consider that the major contribution
to the energy, determining the geometry of a configuration,
comes from the rest energy, i.e. the mass.
Therefore, to make the computation easier, 
instead of the integral of energy gradient we can consider  
the sum of the gradients of the mass distribution:
\be
\int_a
| \nabla_x E |_a ~ \approx ~ \sum_k^{(a)} | \Delta m^{(k)} | \, .
\ee
The ratios of critical temperatures between two such materials should 
approximately be:
\be
{T_c (i) \over T_c (j)} 
~ \sim ~
{\sum_k^{(a_i)} | \Delta m^{(k)}_i  | \over
\sum_{\ell}^{(a_j)} | \Delta m^{(\ell)}_i  |}
\, .
\label{Tnabla}
\ee
This expression will allow us to investigate complex lattice structures.
We stress that what matters is not simply 
the geometric lattice structure,
with geometry intended as the space arrangement of atoms seen as massless
geometrical solids, but the space distribution of energies, 
in the sense of general relativity. 
If in first approximation
we neglect isotope effects, in the purpose of comparing  
ratios of gradients, 
instead of the mass, we may just consider the atomic number. 
We will now apply these considerations to the investigation of
the next step in the table of page~\pageref{tableSc}, Nb$_3$Sn.

\subsubsection{Nb$_3$Sn}

In the case of NbTi, the atomic numbers Nb = 41, Ti = 81 lead to:
\be
\sum | \Delta m| = | 41 - 81 | = 40 ~~~~~~~~
({\rm NbTi}) \, ;  
\ee
for Nb$_3$Sn, Nb = 41, Sn = 50 give:
\be
\sum | \Delta m| = | 41 \times 3 - 50 | = 73 ~~~~~~~~
({\rm Nb}_3{\rm Sn}) \, ;  
\ee
The ratio of the sums of mass gradients of 
${\rm Nb}_3{\rm Sn}$ to $ {\rm NbTi}$ 
is therefore 1,825, that, from \ref{Tnabla} 
and $T_c({\rm NbTi}) \sim 10 ~ ^{\rm o} {\rm K}$ 
should lead to some $18 \div 19$ $^{\rm o}$K 
for the Nb$_3$Sn critical temperature. The observed one is around 
18~$^{\rm o}$K.

\subsection{High temperature superconductors}

Although more complex, high temperature superconductors are 
structured in layers, with a lattice structure that basically develops only
along one coordinate. Their analysis is therefore, in first approximation, 
relatively simple, at least as long as one neglects the doping of 
certain sites with other elements. This introduces a further symmetry breaking
that, in principle, leads to an enhancement of the estimated critical 
temperature. 
This operation may be considered somehow as 
a ``built-in'' ground effect, 
which underlies the properties of any one
of these materials, and as such provides
a systematic error, that can be observed in the general underestimating of the
critical temperature. However, as doping varies from material to
material, this further symmetry 
breaking cannot simply be ``subtracted out'' as a constant, universal
effect: it introduces a further factor of uncertainty and approximation
in our calculations. Our results should therefore be taken more for their
capability to catch the main behaviour, than an attempt to
really provide a fine evaluation of the exact critical temperature. 
As a matter of fact, our estimates fall anyway within an error of at most
$15 \%$ from the experimental observations.

\subsubsection{LaOFeAs and SmOFeAs}

For the iron-based superconductors we consider LaOFeAs and 
SmOFeAs. 
The crystal structure of LaOFeAs is arranged as a stack of
layers in sequence (As) (Fe) (As) (La) (O) (La) etc.
The one of SmOFeAs as a sequence of (As) (Fe) (As) (Sm) (O) (Sm)
(see  \cite{natureLaOFFeAs},
\cite{superc55k} and \cite{NPGsuperc55k}). 
The atomic numbers
La = 57, O = 8, Fe = 26, As = 33 lead to:
\ba
\sum | \Delta m| & = & 
2 |m({\rm As}) - m ({\rm Fe})| + |m({\rm La})-m({\rm As})| 
+ 2 |m({\rm La}) - m({\rm O})| \nn \\
&&+ |m({\rm La})-m({\rm As})| \nn \\
& = & 2 | 33 - 26 | + |57-33| + 2 |57-8| +
|57-33| = 14 + 24 + 98 + 24= 160 \nn \\
&& ~~~~~  ({\rm LaOFeAs}) \, ;   
\label{laofeas}
\ea
Sm = 62, O = 8, Fe = 26, As = 33 lead to:
\ba
\sum | \Delta m|&  = &
2 |m({\rm As}) - m ({\rm Fe})| + |m({\rm Sm})-m({\rm As})| 
+ 2 |m({\rm Sm}) - m({\rm O})| \nn \\
&&+ |m({\rm Sm})-m({\rm As})| \nn \\
& = & 2 | 33 - 26 | + |62-33| + 2 |62-8| + |62 - 33|= 14 + 29 + 108 + 29 
 = 180 \nn \\
&& ~~~~~ ({\rm SmOFeAs}) \, ;   
\ea
This gives as critical temperatures 42 and 47 $^{\rm o}$K respectively.
The observed ones are 44 and 57 $^{\rm o}$K.

\subsubsection{YBCO}

We consider now the yttrium barium calcium copper oxide (YBCO)
\cite{PhysRevLett.58.908}. This 
material superconducts in its orthorhombic form.  
It is arranged as a stack of
layers in sequence (Cu-O) (Ba-O) (Cu-O) (Y) (Cu-O) (Ba-O) (Cu-O).
Differently from
the previous examples, an evaluation of the mass gradients must here
take into account also the fact that not only we have a gradient in passing 
from one layer to the neighbouring one, but also within each of the layers
consisting of bonds of Ba and Cu with oxygen. 
In the planes presenting these
bonds, it is not enough to just consider the gradient with the following 
plane: we must sum up also the mass gradient of the oxygen bond.
On the other hand, 
in order to evaluate the overall gradient to be used in \ref{Tnabla}, 
it is not correct to sum up the absolute 
values of the ``vertical'' and the ``horizontal'' gradient. 
What counts for our purposes is the mean gradient contributed by each plane.
We assume that, as in any
propagation of errors, gradients in the two orthogonal axes sum up 
quadratically, as lengths of orthogonal vectors in a vector lattice. 
The overall gradient should approximately be given by
the sum of the square roots of the quadratically propagated gradients of each 
layer, both in the ``horizontal'' and ``vertical'' directions.
The evaluation of the 
mass gradient is complicated by the fact that, at the transition
to the yttrium layer, oxygen couples both to copper 
and to yttrium, in an orthorhombic form. The crystal is therefore not 
structured in simple layers. 
In order to evaluate the mass gradient for the CuO$_2$-Y 
planes we make the 
approximation of attributing one oxygen atom to the copper layer, and one to 
yttrium.
The expression of the sum of mass gradients is then \footnote{From now on
we adopt the convention of indicating elements with Roman capital letters, 
and in italics their mass, so that e.g. $Cu$ stays for 
$m({\rm Cu})$.}:
\ba
\sum |\Delta m| & = & 2 \times \left\{ \sqrt{[(Cu+O)-(Ba+O)]^2 + (Cu-O)^2}
\right. \nn \\
& & + \, \sqrt{[(Ba+O)-(Cu+O)^2] + (Ba-O)^2} \nn \\
& & + \, \sqrt{[(Cu+O)-O]^2 + (Cu-O)^2} \nn \\
& & + \, \left. \sqrt{(O-Y)^2} \right\} \, . 
\ea
Considering the atomic numbers Y = 39, Ba = 56, 
Cu = 29, O = 8, we have Cu + O = 37, Ba + O = 64, and Cu -- O = 21, 
Ba -- O = 48,
and therefore:
\ba
\sum | \Delta m| & = & 2 \times \left\{ \sqrt{(37 - 64)^2 + 21^2} + 
\sqrt{(64-37)^2 + 48^2} + \sqrt{(37-8)^2 + 21^2} \right. \nn \\
&& \, + \, \left.  \sqrt{(8 - 39)^2} \right\} 
\nn \\ 
& \approx & 2 \times \left\{ 34,2 + 55,1 + 36 + 31 \right\} ~ \approx ~ 312  
~~~~~~~~~~~~~
({\rm YBCO}) \, .   
\label{ybco}
\ea
Rescaling the temperature from the previous elements through \ref{Tnabla},
we obtain a critical temperature 
$T_c \approx 312/160 \times 42 \sim 82$ $^{\rm o}$K.
If, in order to reduce the propagated error,
instead of starting with the critical temperature of LaOFeAs as obtained 
through the series of rescalings from the metallic superconductors, we use as 
starting point its experimental value, 44 $^{\rm o}$K, we obtain for YBCO 
a critical temperature of
$\sim 86$ $^{\rm o}$K. The experimental one is around 
$90 \div 92$ $^{\rm o}$K.

The YBCO compound is part of a series, the so-called ``123'' 
superconductors, of similar critical temperatures, which differ
by the substitution of yttrium with another element of the family
of lanthanoids, including lanthanium.
All these elements are heavier than yttrium, and we expect higher critical 
temperatures. This however is not always what happens. For instance,
(Y$_{0.5}$Gd$_{0.5}$)Ba$_2$Cu$_3$O$_7$  with $T_c$ = 97 $^{\rm o}$K, 
(Y$_{0.5}$Tm$_{0.5}$)Ba$_2$Cu$_3$O$_7$  with
105 $^{\rm o}$K, and (Y$_{0.5}$Lu$_{0.5}$)Ba$_2$Cu$_3$O$_7$ 
with 107 $^{\rm o}$K present an increasing critical 
temperature, as expected from the increasing of mass of the elements that 
substitute the pure yttrium, and the further symmetry breaking due to 
the fact that yttrium is substituted by a mixture of elements, as indicated 
in the brackets.
However, YbBa$_2$Cu$_3$O$_7$ has $T_c$ = 89 $^{\rm o}$K, and 
TmBa$_2$Cu$_3$O$_7$ has $T_c$ = 90 $^{\rm o}$K although Tm is lighter 
than Yb, and similarly  GdBa$_2$Cu$_3$O$_7$ has $T_c$ = 94 $^{\rm o}$K, and 
NdBa$_2$Cu$_3$O$_7$  has 
$T_c$ = 96 $^{\rm o}$K, although Nd is lighter than Gd.
A reason for this apparently odd behaviour could lie in the fact that the
differences in atomic number are indeed very small, to the point that other 
effects play a non negligible role. A finer determination
of the space layout of the energies and masses of these configurations
would be in order.

There is another superconductor very similar to those of the YBCO series.
It is
YSr$_2$Cu$_3$O$_7$, which has a critical 
temperature $T_c$ = 62 $^{\rm o}$K. Strontium 
has atomic number 38, instead of the 56 of barium. In expression \ref{ybco}
we must therefore substitute Ba + O = 64 with 38 + 8 = 46, and Ba + O = 48
with Sr -- O = 30. We have:
\ba
\sum | \Delta m| & = & 2 \times \left\{ \sqrt{(37 - 46)^2 + 21^2} + 
\sqrt{(46-37)^2 + 30^2} + \sqrt{(37-8)^2 + 21^2} \right. \nn \\
&& \, + \, \left.  \sqrt{(8 - 39)^2} \right\}  \nn \\ 
& \approx & 2 \times \left\{ 22,9 + 31,3 + 36 + 31 \right\} ~ \approx ~ 242,4  
~~~~~~~~~~~~~
({\rm YSrCCO}) \, .   
\label{ysrcco}
\ea
Rescaling from YBCO, we obtain
a critical temperature of $63 \div 64$ $^{\rm o}$K, 
in substantial agreement with the experiments.

\subsubsection{BSCCO}

We consider now the bismuth-strontium-calcium-copper-oxide superconductors
(BSCCO) \cite{JJAP.27.L209}: Bi2212 (Bi$_2$Sr$_2$CaCu$_2$O$_2$) and Bi2223 
(Bi$_2$Sr$_2$Ca$_2$Cu$_3$O$_{10}$). The lattice structure of the Bi2212
form is a stack of the following layers: (Bi-O) (Sr-O) (Cu-O$_2$)
(Ca) (Cu-O$_2$) (Sr-O) (Bi-O) (Bi-O) (Sr-O) (Cu-O$_2$)
(Ca) (Cu-O$_2$) (Sr-O) (Bi-O).
The Bi2223 is similar, with one more (Ca) (Cu-O$_2$) layer. 
As for YBCO, here too we must propagate both the ``horizontal'' and the 
``vertical'' gradients. In this case the horizontal bonds are those
of Bi, Sr and Cu with oxygen. 
For the Bi2212 form, we need therefore
BiO = 83+8 = 91, SrO = 38+8=46, Ca = 20, CuO$_2$ = 29+16=45 
and Bi--O = 75, Sr--O  =30, Cu--O = 21, to give:
\ba
\sum | \Delta m| & = & 2 \times \left\{ \sqrt{[(Bi+O)-(Sr+O)]^2+(Bi-O)^2} 
\right. \nn \\
&& + \, \sqrt{[(Cu+2O)-(Sr+O)]^2+(Sr-O)^2} \nn \\
&& + \, \sqrt{[(Ca)-(Cu+2O)]^2+[(Cu-O)+(Cu-O)]^2} \nn \\
& = & 2 \times \left\{ \sqrt{(91 - 46)^2 + 75^2} + 
\sqrt{(45-46)^2 + 30^2} + \sqrt{(20-45)^2 + (21 + 21)^2} \right\} \nn \\ 
& \approx & 2 \times \left\{ 87,5 + 30 + 49 \right\} ~ \approx ~ 332  
~~~~~~~~~~~~~
({\rm Bi2212}) \, .   
\ea
Rescaling the temperature from LaOFeAs through \ref{Tnabla}
we obtain a critical temperature 
$T_c \approx 332/160 \times 42 \sim  87$ $^{\rm o}$K.
Starting from the experimental value, 44 $^{\rm o}$K, in order
to reduce the propagated error, we obtain for the 
Bi2212 a critical temperature of
$\sim 91$ $^{\rm o}$K, closer to the experimental one (92 $^{\rm o}$K).

The structure of Bi2223 is very similar to the one of Bi2212, 
with just the difference of a Ca, CuO$_2$ layer-pair in each
half-lattice block. In order to obtain the mass gradient
of Bi2223 we must therefore just correct the former evaluation
by adding an amount $|m({\rm Ca})-m({\rm CuO}_2)| + 
\sqrt{[m({\rm Ca})-m({\rm CuO}_2)]^2
+ [2|m({\rm Cu})-m({\rm O})|]^2}$:
\ba
\sum | \Delta m| & = & \sum | \Delta m|({\rm Bi}(2212))
+ |20-45| + \sqrt{(20-45)^2 + (21 + 21)^2} \nn \\ 
& = & 332 + 74 ~ = ~ 406  
~~~~~~~~~~~~~~~~~~~
({\rm Bi2223}) \, ,   
\ea
corresponding to a temperature of $406/160 \times 42 = 107$ $^{\rm o}$K
($\sim 111$ $^{\rm o}$K 
if we start from the experimental 44 $^{\rm o}$K for the critical 
temperature of LaOFeAs). The experimental value is around 110 $^{\rm o}$K.

\subsection{The Tl-Ba-Ca-Cu-O superconductor}

\subsubsection{Tl$_2$Ba$_2$CuO$_6$ (Tl-2201)}

The stacking sequence is as follows: 
(Tl-O) (Ba-O) (Cu-O$_2$) (Ba-O) (Tl-O) \footnote{See \cite{tlsupc},
\cite{tlseries}, and also~\cite{khare}.}, and
the expression of the mass gradient sum is:
\ba
\sum | \Delta m| & = & \sqrt{(Tl-O)^2 + [(Tl+O)-(Ba+O)]^2} \nn \\
& + & \sqrt{[(Ba+O)-(Cu+O+O)]^2 + (Ba-O)^2} \nn \\
& + & \sqrt{(Cu-O)^2+(Cu-O)^2 + [(Cu + O + O)-(Ba + O)]^2} \nn \\
& + & \sqrt{[(Ba+O)-(Tl+O)]^2 + (Ba - O)^2} \nn \\
& + & \sqrt{(Tl - O)^2 + [(Tl+O)-(Tl+O)]^2} \, .
\ea
From the atomic numbers Tl = 81, Ba = 56, Cu = 29 and O = 8 we derive  
(Tl-O) = 73, (Tl + O) = 89, (Ba--O) = 48,
(Ba + O) = 64, (Cu--O--O) = 21, and (Cu+O+O) = 45.
Plugging these values into the gradient sum expression, we obtain:
\ba
\sum | \Delta m| & = & \sqrt{73^2 + (89-64)^2} \nn \\
& + & \sqrt{(64-45)^2 + 48^2 } \nn \\
& + & \sqrt{2\times 21^2 + (45-64)^2} \nn \\
& + & \sqrt{(64-89)^2 + 48^2} \, + \, 73 \nn \\
& = & 291 \, .
\ea 
Rescaling now from LaOFeAs, expression \ref{laofeas},
and using once again the 44 $^{\rm o}$K of the experimental temperature,
we obtain $(291/160) \times 44 = 80$ $^{\rm o}$K (had we used our calculated
42 $^{\rm o}$K for LaOFeAs, we would have obtained $\sim 76,5$ $^{\rm o}$K). 
The experimental critical 
temperature is around 80 $^{\rm o}$K.

\subsubsection{Tl$_2$Ba$_2$CaCu$_2$O$_8$ (Tl-2212)}

In this crystal there are two Cu-O-O layers with a Ca layer in between,
with stacking sequence (Tl-O) (Ba-O) (Cu-O$_2$) (Ca) (Cu-O$_2$) (Ba-O) (Tl-O).
In order to obtain the mass-gradient sum we have just to add to the
previous computation a module:
\be
\sqrt{[(Cu+O+O)-Ca]^2 + (Cu-O)^2 + (Cu-O)^2} \, + \, 
\sqrt{[(Cu+O+O)-Ca]^2} \, . 
\ee
Considering that the atomic number of Ca is 20, this means an amount:
\be
\sqrt{25^2 + 2 \times 21^2} \, + \, 25 ~ = ~ 64 \, .
\ee
This gives a sum $291 + 64 = 355$, leading to a critical temperature of 
around 98 $^{\rm o}$K.
The experimental one is around 108 $^{\rm o}$K.

\subsubsection{Tl$_2$Ba$_2$Ca$_2$Cu$_3$O$_{10}$ (Tl-2223)}

In this crystal
there are three CuO$_2$ layers enclosing one Ca layer between each of them.
That means, one more [(CU-O-O) Ca] module as compared to 
Tl$_2$Ba$_2$CaCu$_2$O$_8$.
We obtain therefore a value of mass gradient sum 
$355 + 64 = 419$, leading to a critical temperature
of 115 $^{\rm o}$K. The experimental one is 125 $^{\rm o}$K.

\
\\

Both in this and in the previous superconductor we obtain slightly 
underestimated values
of critical temperature. On the other hand, the ratio of the two critical 
temperatures we obtain, namely,
$115/98$, is in better agreement with the ratio of the
experimental values. 
Indeed, it gives a slight overestimate, 
which partially
compensates the underestimate of the first temperature.
From a qualitative point of view, these under/over-estimates can be understood 
as follows: when a Ca layer is added to the Tl$_2$Ba$_2$CuO$_6$ structure,
the symmetry of the configuration of a stack of ``(X-O)'' layers 
gets further broken, because the Ca layer does
not contain an oxygen bond. Not taking this into account leads to an 
underestimate
of the increase in critical temperature. On the other hand, when a further
identical layer is added, there is a partial restoration of symmetry, which
implies a reduction in the increase of critical temperature, thereby our 
over-estimation.
This effect becomes more relevant in more complicated configurations:
in Tl-based superconductors, the value of $T_c$  decreases after four CuO$_2$ 
layers in TlBa$_2$Ca$_{n-1}$Cu$_n$O$_{2n+3}$, 
and in the Tl$_2$Ba$_2$Ca$_{n-1}$Cu$_n$O$_{2n+4}$  compound it decreases 
after three CuO$_2$ layers \cite{tlsupc2}.

\subsection{Comparing within families}

Superconductivity is detected through investigation
of the magnetic properties of materials. 
In particular, for what concerns high-temperature
superconductors, pinning of magnetic flux through impurities 
plays a significant role,
not only in reducing the effective resistance, 
and therefore affecting the conditions for the detection of a regime
recognizable as the one of superconductivity, but, in the light 
of our analysis, also because it decreases the symmetry of the configuration.
Also pressure plays a relevant role, because high pressures correspond to more 
remote configurations, and are expected to lead to higher critical 
temperatures (a fact that
corresponds to the experimental observation). It is therefore rather difficult
to give a correct quantitative account of the superconducting properties 
and the critical temperatures of all
superconducting materials, and impossible to do it only in terms of 
comparison of average mass gradients referring to 
a single material taken as a universal starting point.
In several cases, the best we can do is comparing critical temperatures within
``families'' of materials, which are assumed to share common properties, so
that the change in the lattice structure taken into account by our evaluation
of mass gradients can be considered as 
the only relevant variable and effective term of comparison.

\subsubsection{Hg-Ba-Ca-Cu-O superconductor}

An example of this kind of difficulties is provided by the Hg-series (Hg-1201,
Hg-1212, Hg-1223 ~\cite{putilin}). 
In principle, it is analogous to the series in which
mercury is substituted by thallium (the Tl-series: 
Tl-1201, Tl, 1212, Tl-1223), but, while the critical 
temperature of Tl-1201 is lower than 10 $^{\rm o}$K, 
the one of the analogous compound
made with Hg (one position lower in the atomic number scale) 
is around 94 $^{\rm o}$K.
Both these numbers escape the predictions we can make with our simple
mass-gradient arguments, applied using mercury as starting point. 
Indeed the Hg-1201 material is a critical example in which 
doping plays crucial role, whose details are still controversial.
As reported in \cite{hgcontrov}, depending on the amount of doping, this
cuprate can superconduct or not, with a range
of critical temperatures spanning the whole spectrum from zero to the
maximal value. The critical temperature has proven to be also very sensitive 
to pressure \cite{hgpressure}. 
In this case, the best we can do is to compare critical temperatures
assuming comparable doping/flux pinning conditions.
Assuming that, for instance, the highest critical temperature
within the Hg-1201, 1212, 1223 series are obtained with
a similar amount of such ``external'' inputs,
we can expect to be able to give a reasonably good
estimate of the ratios 
of critical temperatures \emph{within} the Hg series.
An illustration
of the crystal structure of HgBa$_2$CuO$_4$ (Hg-1201, $T_c$ = 94 $^{\rm o}$K),
HgBa$_2$CaCu$_2$O$_6$ (Hg-1212, $T_c$ = 128 $^{\rm o}$K) and 
HgBa$_2$Ca$_2$Cu$_3$O$_8$  (Hg-1223, $T_c$ = 134 $^{\rm o}$K) 
can be found in \cite{khare}. 
Computing the ratios of temperatures along the same line as in the previous
examples, we obtain
\ba {T_c (Hg-1223) \over T_c (Hg-1212)} & \sim & 1,23 \, , \nn \\
{T_c (Hg-1212) \over T_c (Hg-1201)} & \sim & 1,3 \, , \nn \\ 
\left( {T_c (Hg-1223) \over T_c (Hg-1201)} \right.
& \sim & \left. 1,59 
\hspace{-.3cm}\begin{array}{c} \mbox{} \\ \mbox{} \end{array} \right) 
\, , 
\ea
to be compared with
the ratios of the experimental ones, namely $1,05$, $1,36$, and $1,43$ for the
($1223 \big/ 1201$) ratio.
They show a similar situation of underestimate for the
ratio of the lower pair of temperatures, 
and overestimate for the ratio of the third to 
the second one, as in the case of the thallium compound discussed above.
Taking this into account, the ratios we find are not far from the 
experimental ones
(the absolute determination of the temperature fails in this case
to give a correct prediction, in that it
would tell that both the thallium and the mercury -1201, -1212, -1223 series
should have the same critical temperatures).  
This suggests that, keeping fixed all other conditions, the argument
based on the evaluation of symmetry properties of the mass/energy 
configurations
makes sense, although in some cases it is too simplified, and not sufficient to
determine the overall conditions producing the particular state
of a material which is detected as a regime of superconductivity.

\vspace{1cm}

A comparison restricted to elements belonging to the same family 
is our way of proceeding also in the case of higher temperature
superconductors.
Indeed, when passing to higher-$T_c$ superconductors, and therefore to
higher complexity of the lattice structure, a thorough analysis of the details
of any part of the lattice block becomes the more and more difficult.
On the other hand, in first approximation a detailed knowledge
of the full lattice structure is not even necessary. 
The materials we are going
to consider can be grouped in ``families'', 
whose elements share part of the lattice structure, and
differ by the structure of just one (or some) of the lattice
blocks. In this way, it is possible to perform a partial analysis, by comparing
the critical temperatures among the members of each family.  
As the whole structure becomes longer and longer, 
it becomes smaller the error we introduce in weighting the various blocks
according to their average length, thereby neglecting the details of the single
mass gradients within common blocks.
The difference from one material to the neighbouring one 
within a family usually consists in the substitution of some atomic elements,
or in the addition of further replicas of already present layers.
The mass differences introduced by these changes  
will be dealt with as a ``second order'' perturbation:
\be
{T^{\prime}_c  \over T_c} ~ = ~ {T_c 
+ \delta T_c \over {\rm T}_c} ~ = ~
1 + {\delta T_c \over T_c }  ~ \approx ~
1 + \delta |(\nabla M)|_{\rm extra~block} \Big/ \sum |\nabla M| \, .  
\ee

\subsubsection{The SnBaCaCuO to (TlBa)BaCaCuO family.}

\begin{itemize}

\item{\sl From {\bf 160 $^{\rm o}$K} (Sn$_3$Ba$_4$Ca$_2$Cu$_7$O$_{\nu}$)
to {\bf 200 $^{\rm o}$K} (Sn$_6$Ba$_4$Ca$_2$Cu$_{10}$O$_{\nu}$).}

The lattice structure of Sn$_3$Ba$_4$Ca$_2$Cu$_7$O$_y$ consists of a stack
of (Ca) (CuO$_x$) 
[ (Ba) (CuO$_y$) (Ba) ] (CuO$_x$) (Ca) (CuO$_x$) 
[ (Ba) (Sn-O) (Cu) (Sn-O) (Cu)
(Sn-O) (Ba) ] (CuO$_x$) , where in the first square bracket we indicate
the light part of the lattice, in the second the heavy part,
and (CuO$_x$), (CuO$_y$) indicate copper oxide layers. Here and in the 
following we
use this notation to indicate, in general, (CuO$_3$) and (CuO$_2$)
layers respectively \footnote{Illustrations of this structure and
of those of the following materials can be found in the Joe Eck's website,
www.superconductors.org.}.   
The lattice structure of 
Sn$_6$Ba$_4$Ca$_2$Cu$_{10}$O$_{\nu}$ is obtained by
doubling the ``heavy'' part of the 
lattice of Sn$_3$Ba$_4$Ca$_2$Cu$_7$O$_{\nu}$.
The structure of this superconductor corresponds therefore to that 
of Sn$_3$Ba$_4$Ca$_2$Cu$_7$O$_{\nu}$,
with the duplication of an entire lattice block. For the evaluation of the 
critical temperature we assume that, owing to the high number of lattice
elements/layers, in first approximation we can consider the geometry of the 
blocks structure as prevailing over the fine structure of energy gradients, 
which distinguishes between light and heavy part of the lattice. That means,
in first instance we deal with the blocks as if all lattice layers were equal, 
something that in the average is not far from the truth, and implies an error
that becomes smaller and smaller as we go on with an increasing length of 
the crystal structure. 
Since the ``replica'' of the lattice block we add to obtain this superconductor
corresponds to around 1/4  of 
the whole structure, we expect some 25\% of increase in $T_c$ 
from the one above, corresponding to an increase from 
160 to 200 $^{\rm o}$K \footnote{More precisely, 
since the change is made in the heavy part of the lattice,
it corresponds to more than 1/4 of the structure. However, since the
modification consists in adding a replica of one layer, the effect
is softened by the fact that there is also a further symmetry among the two
identical layers. 1/4 is therefore to be taken as a rough estimate of the
order of magnitude of the effect.}.

\item{\it  {\bf 212 $^{\rm o}$K}: (Sn$_5$In)Ba$_4$Ca$_2$Cu$_{10}$O$_{\nu}$}

The lattice structure consists of a stack of the following
layers: 
(Ca) (CuO$_x$) 
[ (Ba) (CuO$_2$) (Ba) ] (CuO$_x$) (Ca) (CuO$_x$) 
[ (Ba) (Sn,In-O) (Cu) (Sn,In-O) 
(Cu) (Sn,In-O) (Cu) (Sn,In-O) (Cu) (Sn,In-O) 
(Cu) (Sn,In-O) (Ba) ] (CuO$_x$).
We expect a higher $T_c$ than the crystal of above (200 $^{\rm o}$K), 
as a consequence of the lower symmetry, now broken by the substitution of
a tin atom with indium. This corresponds 
to a breaking of more or less one out of $18 \div 20$ lattice layers, i.e. a 
$\sim 5 \div 6 \%$ of the total. Since 
the mass difference between Sn and In is of
much lower order, in first approximation  
the increase of the critical temperature should 
be mainly determined by the symmetry breaking among different
lattice layers, and therefore be of order $\sim 5 \div 6 \%$. 
This gives indeed some 210-212 $^{\rm o}$K, as is observed.

The mass difference between Sn and In plays a role as a second order effect,
that can be observed in the smaller variation of the critical 
temperature after a change of the (Sn$_5$In) structure 
into (Sn$_4$In$_2$). The (Sn$_5$In)
compound should superconduct at a higher temperature than (Sn$_4$In$_2$),
where there is a partial reconstruction of a higher symmetry within
indium planes.
Experimentally, one observes 212 $^{\rm o}$K for the first, 
and 208 $^{\rm o}$K for the second. 
Also in this case, an evaluation, even approximate, is rather difficult, 
because the naive value of 5/4 one would suppose (20\% increase in the 
temperature) must be "tempered" by the fact that Sn and In weight almost the 
same. Their relative mass difference is 1/50, and this would mean a symmetry 
breaking of about 2\%, indeed corresponding to the order of change in 
the observed $T_c$.

\item{\it  {\bf 218 $^{\rm o}$K}: (Sn$_5$In)Ba$_4$Ca$_2$Cu$_{11}$O$_{\nu}$}

The lattice structure is similar to the one of 
(Sn$_5$In)Ba$_4$Ca$_2$Cu$_{10}$O$_{\nu}$ but contains one extra Cu in the
light part of the lattice:
(Ca) (CuO$_x$) 
[ (Ba) (Cu$_2$O$_y$) (Ba) ] (CuO$_x$) (Ca) (CuO$_x$) 
[ (Ba) (Sn,In-O) (Cu) (Sn,In-O) 
(Cu) (Sn,In-O) (Cu) (Sn,In-O) (Cu) (Sn,In-O) 
(Cu) (Sn,In-O) (Ba) ] (CuO$_x$). Adding a 
copper atom breaks part of the symmetry, thereby increasing $T_c$. 
As this occurs 
in one of the some 22 lattice "stairs", we would expect this to produce a 
correction of about $\sim$1/22 = 4,5\% of $T_c$. 
This would mean some 9 $^{\rm o}$K. However, in this estimate we don't consider
finer corrections obtained by taking into account
mass gradients. In practice, the breaking of symmetry is softened
by the fact that there is a partial restoration of symmetry due to the 
fact that we are adding one more atom in a layer made
of atoms of the same element. Indeed,
the correction which is experimentally observed seems to be 
around 6 $^{\rm o}$K,
indicating a slightly lower symmetry breaking than in our rough estimate.

\item{\it  {\bf 233 $^{\rm o}$K}: Tl$_5$Ba$_4$Ca$_2$Cu$_{11}$O$_{\nu}$}

The lattice structure is given by the following stack:
(Ca) (CuO$_x$) [ (Ba) (Cu$_2$O$_y$) (Ba)] 
(CuO$_x$) (Ca) (CuO$_x$) [ (Ba) (Tl-O) (Cu) (Tl-O) 
(Cu) (Tl-O) (Cu) (Tl-O) (Cu) (Tl-O) (Ba) ] (CuO$_x$).
The heavy part of the lattice is similar to the one of the previous cuprate,
with the suppression of the indium layer, and the substitution 
of tin atoms with thallium. In order
to compare critical temperatures,
let us compute the mass gradient sums corresponding
to this part of the lattice for both these
materials. They correspond to the stacking sequences  
(Ba) (Sn-O) (CuO) (Sn-O) (Cu) (Sn-O) (Cu) (Sn-O) (Cu) (Sn-O) (Cu) (Sn-O) (Ba)
and (Ba) (Tl-O) (Cu) (Tl-O) (Cu) (Tl-O) (Cu) (Tl-O) (Cu) (Tl-O) (Ba)
respectively. In the Sn sequence, one of the six tin atoms is substituted
by an indium atom. Since the atomic numbers are respectively
In = 49 and Sn = 50, in first approximation we neglect the slight asymmetry
introduced by the (5 Sn)/In alternance.
The mass gradient sums are:
\ba
\sum | \Delta m | & = & |Ba - (Tl + O) | \nn \\
&& \, + \, 4 \times \left\{ \sqrt{(Tl - O)^2 + [(Tl + O) - Cu]^2} \, + \,
|Cu - (Tl + O)| \right\} \nn \\
&& \, + \, \sqrt{(Tl - O)^2 + [(Tl + O) - Ba]^2} \, ,
\ea
and, neglecting the difference between Sn and In:
\ba
\sum | \Delta m | & = & |Ba - (Sn + O) | \nn \\
&& \, + \, 5 \times \left\{ \sqrt{(Sn - O)^2 + [(Sn + O) - Cu]^2} \, + \,
|Cu - (Sn + O)| \right\} \nn \\
&& \, + \, \sqrt{(Sn - O)^2 + [(Sn + O) - Ba]^2} \, .
\ea
Inserting the atomic numbers Ba = 56, Tl = 81, Cu = 29, O = 8 and Sn = 50
we obtain:
\ba
\sum | \Delta m| ({\rm Tl}_5 ) & = &
|56 - 89| \, + \, 
4 \left\{ \times \sqrt{73^2 + (89-29)^2} \, + \, |29-89| \right\} \nn \\
&& \, + \, \sqrt{73^2 + (89 - 56)^2} \nn \\
& = & 
\, 33 + 4 \times \left\{ 94,5 + 60  \right\}  + 80,1 
~ \approx ~ 731 \, ,
\ea
and 
\ba
\sum | \Delta m| ({\rm Sn}_5 {\rm In}) & = &
|56 - 58| \, + \, 
5 \left\{ \times \sqrt{42^2 + (58-29)^2} \, + \, |29-58| \right\} \nn \\
&& \, + \, \sqrt{42^2 + (58 - 56)^2} \nn \\
& = & 
\, 2 + 5 \times \left\{ 51 + 29  \right\}  + 42 
~ \approx ~ 444 \, .
\ea
The ratio between the two sums is therefore:
\be
{\sum |\Delta m|({\rm Tl}_5 ) \over  \sum | \Delta m| ({\rm Sn}_5 {\rm In}) }
~ \approx ~ 1,65 \, .
\ee
In order to derive the rescaling of the critical temperature, we must see
how much these gradients weight in the overall determination
of the symmetry of these crystal configurations. The heavy part of the lattice 
amounts to more or less one half of the entire structure.
However, a large part of this sub-lattice has a symmetry of five--almost six 
layers respectively. In practice, if the gradient (Ba)--(Tl-O), or
(Ba)--(Sn-O) occurs on two stairs out of some 20--22, the change
from (Sn-O)--(Cu) to (Tl-O)--(Cu), while occurring along some 5 layers,
does not contribute so much to the reduction of symmetry. Owing to
the symmetry of this stack, we expect it to contribute only by a factor
$\sim {1 \over 5!}$. Within the order of approximation we are making in this
evaluation, it can therefore be neglected.
The only part that counts is therefore the ratio:
\be
{\sqrt{[Ba - (Sn + O)]^2 + (Sn - O)^2} \over
\sqrt{[Ba - (Tl + O)]^2 + (Tl - O)^2}} ~ \approx ~
1,9 \, ,
\ee
that corresponds to the change in two out of some 20 layers, giving
therefore a factor:
\be
{\langle || G || \rangle_{{\rm Tl}_5} \over 
\langle || G || \rangle_{{\rm Sn}_5 {\rm In}}  }
~ \approx ~ {1,9 + 10 \over 11} ~ \sim ~ 1,082 \, ,
\ee
implying a jump in critical temperature from 218 $^{\rm o}$K to 
236 $^{\rm o}$K.

\item{\it  {\bf 242 $^{\rm o}$K:} (Tl$_4$Ba)Ba$_4$Ca$_2$Cu$_{11}$O$_{\nu}$}

The lattice structure is a stack of the following layers:
(Ca) (CuO$_x$) [ (Ba) (Cu$_2$O$_y$) (Ba) ] 
(CuO$_x$) (Ca) (CuO$_x$) [ (Ba) (X$_1$-O) (Cu) (X$_2$-O) 
(Cu) (X$_3$-O) (Cu) (X$_4$-O) (Cu) (X$_5$-O) (Ba) ] (Cu),
where, for every column, $X_i$ stays four times for Tl, and one time
for Ba in always different position for every layer.
Between the cuprate of above and this one 
there is the substitution of some atoms of 
thallium with barium, which breaks part of the symmetry of 
the heavy part of the lattice. 
In this case, owing to the alternating position of the barium
substituting thallium, the breaking
of the symmetry,
no more negligible as was the case of the Sn/In asymmetry, occurs 
not only in the ``vertical'' but also in the ``horizontal'' direction.
In the aim of estimating the amount of symmetry breaking,
we can make the approximation of considering 
just the effect of neighbouring lattice sites. In this approximation,
each oxygen is surrounded either by four thallium, or by
three thallium and one barium atom. The first case occurs only on one stair,
whereas the other case occurs in the four remnant stairs. 
Therefore, we can roughly say that 
of the initial five thallium layers, four get separated into 
1 (barium) plus 3 (thallium). In each of these four the symmetry factor
is therefore ${2 \over 3}$ (the barium/thallium mass ratio) $\times$
${4 \over 3}$ (the amount of remnant symmetry group, i.e. the ratio of the four
of before the breaking to the three after the breaking).
All in all this makes:
\be
{5 \over 1 + 4 \left( {2 \over 3} \times { 4 \over 3 }  \right)} ~ \approx
~ 1,09756 \, .
\ee
Made on around 1/3 of the whole lattice raw, this implies a jump in the 
critical temperature of a factor around $(2 + 1,09756)/3$, 
that is,
from the former 233-234 $^{\rm o}$K to some 241-242 $^{\rm o}$K.

\item{\it  {\bf 254 $^{\rm o}$K:} (Tl$_4$Ba)Ba$_2$Ca$_2$Cu$_7$O$_{13+}$}

The lattice structure consists of a stack of
(Ca) (CuO$_y$) (Ca) (CuO$_x$) [ (Ba) (X$_1$-O) (Cu) (X$_2$-O) 
(Cu) (X$_3$-O) (Cu) (X$_4$-O) (Cu) (X$_5$-O) (Ba) ] (CuO$_x$).
As compared to the previous one, here the light part of the lattice has been 
partly cut out. 
In this case, differently from what one could expect, shortening 
a piece of the crystal structure leads to an increase of critical
temperature. This can be understood as follows.
All the elements of this family of materials are characterized by
the fact of having a lattice structure composed
of a heavy and a light part. When considered from the point of view of a
larger scale than just one lattice period, the reduction of the
part with lighter masses, although in itself leading to
a lower overall mass gradient within the single lattice length,
on a scale of several lattice units,
owing to the shorter light-lattice structure, it increases 
the average gradient. The average effect is therefore
equivalent to an increase of the heavy part of the lattice, the one 
with higher mass gradients. These situations are illustrated in 
figure~\ref{bigsmallgrad}.
\begin{figure}
\centerline{
\epsfxsize=8cm
\epsfbox{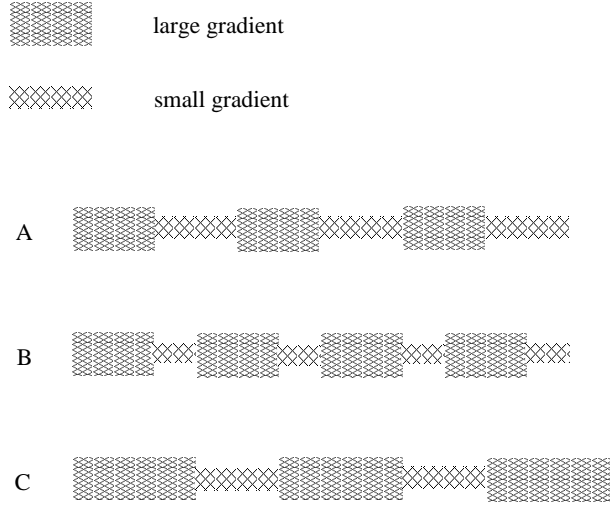}
}
\vspace{0.3cm}
\caption{Both shortening the light, small mass gradient part 
(example B), 
and lengthening the heavy, 
high gradient part of the lattice (example C)
lead to an effective increase of average mass gradient as compared to A, 
and therefore to
a higher remoteness of the configuration, which reflects in an increased
critical temperature.}      
\label{bigsmallgrad}
\end{figure}
We can give a rough estimate of the effect, by considering that
the light part has masses which are around one-half of those of the 
heavy part, and the change in the structure, as
compared to the longer lattice form 
(the one of (Tl$_4$Ba)Ba$_4$Ca$_2$Cu$_{11}$O$_{\nu}$), amounts to suppressing
some 2-3 layers in this light part, out of a total of $\sim$ 20 lattice
planes. This is a change of around 1/2 of $10 \%$, i.e. $\sim 5 \%$,
corresponding to a jump in the temperature of some 12 $^{\rm o}$K.
This leads from the former 242 $^{\rm o}$K to around 254~$^{\rm o}$K.
  
This example shows that, 
although working within a single unit of lattice length,
as implied in \ref{nablaE}--\ref{Tnabla},
is in most cases correct, in principle the comparison of geometries
is something more subtle. In the case of 
(Tl$_4$Ba)Ba$_2$Ca$_2$Cu$_7$O$_{13+}$,
just considering one unit of lattice length is not enough.

\end{itemize}

\subsubsection{The (SnPbIn)BaTmCuO family: from 163 $^{\rm o}$K to 195 $^{\rm o}$K.}

The lattice structure of 
(Sn$_{1.0}$Pb$_{0.5}$In$_{0.5}$)Ba$_4$Tm$_4$Cu$_{6}$O$_{18+}$, 
$T_c$ = 163 $^{\rm o}$K consists of a stack of
(0,5(Sn$_{1.0}$Pb$_{0.5}$In$_{0.5}$)-O) (Ba) (CuO$_x$) (Tm) (CuO$_x$)
(Ba) (0,5(Sn$_{1.0}$Pb$_{0.5}$In$_{0.5}$)-O) (Ba) (CuO$_x$)
(Tm) (CuO$_y$) (Tm) (CuO$_y$) (Tm) (CuO$_x$) (Ba). 
\newline The lattice structure of 
(Sn$_{1.0}$Pb$_{0.5}$In$_{0.5}$)Ba$_4$Tm$_5$Cu$_{7}$O$_{20+}$,
$T_c$ = 185 $^{\rm o}$K consists of a stack of
(0,5(Sn$_{1.0}$Pb$_{0.5}$In$_{0.5}$)-O) (Ba) (CuO$_x$) (Tm) (CuO$_x$)
(Ba) (0,5(Sn$_{1.0}$Pb$_{0.5}$In$_{0.5}$)-O) (Ba) (CuO$_x$)
(Tm) (CuO$_y$) (Tm) (CuO$_y$) (Tm) (CuO$_y$) (Tm) (CuO$_x$) (Ba).
The lattice structure of 
(Sn$_{1.0}$Pb$_{0.5}$In$_{0.5}$)Ba$_4$Tm$_6$Cu$_{8}$O$_{22+}$, 
$T_c$ = 195 $^{\rm o}$K, consists of a stack of
(0,5(Sn$_{1.0}$Pb$_{0.5}$In$_{0.5}$)-O) (Ba) (CuO$_x$) (Tm) (CuO$_x$)
(Ba) (0,5(Sn$_{1.0}$Pb$_{0.5}$In$_{0.5}$)-O) (Ba) (CuO$_x$)
(Tm) (CuO$_y$) (Tm) (CuO$_y$) (Tm) (CuO$_y$) (Tm) (CuO$_y$) (Tm)
(CuO$_x$) (Ba).
It is difficult to compare with the elements of the previous series. 
On the other hand, since the differences among the elements of this series 
consist in adding a [(CuO$_y$) (Tm)] pair of layers 
within the same lattice subset, 
it is relatively easy to compare the 
elements within this group. In practice we are adding a Tm line 
at each step,
increasing the lattice complexity by one layer out of a total of 10 in the 
first case, and of 10+1 in the second case. We expect therefore an increase of 
$T_c$ by a factor
11/10 and 12/11 respectively. This corresponds to a jump from $\sim$163 to 
$\sim$180~$^{\rm o}$K, and to $\sim$195 $^{\rm o}$K in the second case.

\section{Comments}

The analysis of the previous section provides support to the hypothesis
that quantum gravity effects may be at the ground of the understanding
of the relation between lattice complexity and critical temperature
of superconductors.
Roughly speaking, 
working in a quantum gravity framework effectively means 
having a Planck constant dependent on the distribution of energy along space.
If we introduce an energy density $\rho(E)$,
this in practice means that we are effectively promoting $\hbar$ to 
\be
\hbar ~ \to ~ \hbar (\rho(E)) \, .
\ee
Alternatively, since energy distribution along space and
space geometry are equivalent, we can also say that we work
with a geometry-dependent Planck constant:
\be
\hbar ~ \to ~ \hbar (g_{\mu \nu}) \, ,
\ee
or, to work with quantities independent on the choice of
coordinate system, with a curvature-dependent Planck constant.
Although all the expressions considered in this paper are worked out
within a non-field theoretical framework, from an heuristic point of view
this dependence can be understood as follows.
Quantization of gravity introduces an effective dependence on $\hbar$ in
the modes of propagation of the metric tensor $g_{\mu \nu}$. This means that,
even if we start with a space with a classical background metric,
after quantization, and as a consequence of the back reaction
due to the interaction with matter and radiation,
we will end up with a space with $\hbar$-dependent geometry,
$g_{\mu \nu} (\hbar)$. Taking the point of view of 
considering geometry as a primary, independent input corresponds to
inverting the relation $g_{\mu \nu} = g_{\mu \nu} (\hbar)$ to
$\hbar = \hbar (g_{\mu \nu})$.
The functional dependence is not simple;
on the other hand, its explicit expression is not
even fundamental, because it expresses only an effective parametrization:
in general, in order to derive, case by case,
the appropriate effective parametrization, one has
to refer to \ref{Zsum} .
The ground value of the Planck constant is the one corresponding to the 
``vacuum'', which in our case is the universe with uniform curvature,
corresponding to the cosmological constant \footnote{More precisely,
the ground curvature is the average sum of the cosmological term, plus
the contributions of matter and radiation. It is the sum of all these terms 
what gives the universe the ground average curvature of a three sphere
(see Ref.~\cite{spi}). }. A uniform curvature gives a universal contribution
that can be subtracted, i.e. re-absorbed into a redefinition
of the Planck constant. This is what is done when gravity is decoupled
from the quantum theory, and one recovers the traditional quantum theory.

A dependence of the Planck constant on the geometry means that
also the amount of quantum delocalization of wave functions, 
the mechanism at the ground
of superconductivity, depends on the geometry. However, the relation between
critical temperature and lattice complexity of superconductors
cannot be observed in a clean, direct way: superconductivity is a regime
in most cases ``unstable'' in pure materials, and the way it is detected makes
measurements very sensitive to several additional conditions. 
The simple relation we suggest between critical temperature 
and ``geometry'' 
only works at the net of
any other effect, such as degree of doping/pinning of magnetic flux, etc. 
A quantitative prediction is only possible when 
the contribution of these effects can been subtracted. 
The agreement between expectations and
theoretical predictions we find 
has therefore to be read ``in the average'', and works 
better in comparing temperatures between materials belonging to
the same ``family'', for which therefore other conditions can be assumed to
be similar (the case of the Hg-12010/1212/1223 series is exemplar
of this situation). 
Nevertheless, the agreement between predictions and
experimental observations is impressive.
Our analysis provides a further indication that, differently from
what one is used to expect, quantum gravity is not just a matter of 
Planck scale phenomena, but in principle comes into play, to contribute 
for non-negligible corrections, in any
quantum system corresponding to a 
non-trivial geometry of space-time.

\vspace{.5cm}

In the universe described in Ref.~\cite{assiom},
geometry evolves with time. Not only the cosmological
constant and the matter density of the universe, and therefore
the ground curvature, evolve with time,
but also all elementary masses depend on the age of
the universe.
It is not hard to realize that, since
also couplings, in particular the fine structure constant, evolve with the age
of the universe, also the masses of atoms and molecules do evolve. 
In particular, electron, proton and neutron mass evolve
as appropriate inverse powers of the age of the universe:
\[ 
m_i \sim {\cal T}^{- a_i} \, ,
\]
where $a_i < 1$ and $i$ indicates 
the electron, or the 
proton, or the neutron. Similarly it goes for the electric coupling:
\[
\alpha \sim {\cal T}^{- a_{\alpha}} \, .
\]
As a consequence, also the masses of atoms and molecules
are expected to scale with time, perhaps with a more complicated
dependence than that of elementary particles, but anyway in such a way to
decrease with time, with an increasing relative rate. This means that
the mass ratio of heavier to lighter atoms is expected to increase with time.
Therefore, owing to the change of geometry, or, equivalently, of the
effective Planck constant,
we should expect also an increase in the 
ratio of different degrees of delocalization of wavefunctions.
How fast should this go can be estimated by 
considering that, approximately, mass ratios 
scale as powers of the age of the universe.
With a similar degree of approximation, we can assume that also
lattice gradient ratios scale as powers of the age of the universe.
A factor 2 in the ratio of the
mean weights of configurations at present time:
\[
{\xi_i \over \xi_j} ~ \approx ~ {\langle \nabla m_i
\rangle \over \langle \nabla m_j  \rangle} ~  \sim ~ 2 \, ,
\]
corresponds
to a very small exponent $a_{(\xi_i / \xi_j)}$ of the evolution:
\[
{\xi_i \over \xi_j} ~ \approx ~ {\cal T}^{a_{(\xi_i / \xi_j)}} \, .
\]
This is given in fact as 
$\log 2 = a_{(\xi_i / \xi_j)} 
\log {\cal T}$, where the age of the universe ${\cal T}$ 
is expressed in units of 
appropriately converted Planck length.
At present time ${\cal T} \sim 10^{61}$. 
This kind of evolution is therefore only detectable on a large,
cosmological, time scale, and negligible for usual purposes.

\vspace{1.5cm}

\providecommand{\href}[2]{#2}\begingroup\raggedright\endgroup

\newpage

\noindent
\begin{tabular}{| c | c | c |}
\hline
Transition &  & \\
Temperature & Material & Class \\
in Kelvin & & \\
\hline
254 & (Tl$_4$ Ba) Ba$_2$ Ca$_2$ Cu$_7$ O$_{13+}$ & \\
\cline{1-2}
242 & (Tl$_4$ Ba) Ba$_4$ Ca$_2$ Cu$_{11}$ O$_{\nu}$  & \\
\cline{1-2}
233 & Tl$_5$ Ba$_4$ Ca$_2$ Cu$_{11}$ O$_{\nu}$ 
& \\
\cline{1-2}
218 & (Sn$_5$ In) Ba$_4$ Ca$_2$ Cu$_{11}$ O$_{\nu}$ & \\
\cline{1-2}
212 & (Sn$_5$ In) Ba$_4$ Ca$_2$ Cu$_{10}$ O$_{\nu}$ & \\
\cline{1-2}
200 & Sn$_6$ Ba$_4$ Ca$_2$ Cu$_{10}$ O$_{\nu}$ & \\
\cline{1-2}
160 & Sn$_3$ Ba$_4$ Ca$_2$ Cu$_{7}$ O$_{\nu}$ & \\
\cline{1-2}
& & \\
\cline{1-2}
195 & (Sn$_{1.0}$ Pb$_{0.5}$ In$_{0.5}$)Ba$_4$Tm$_6$Cu$_{8}$O$_{22+}$ & \\
\cline{1-2}
185 & (Sn$_{1.0}$ Pb$_{0.5}$ In$_{0.5}$)Ba$_4$Tm$_5$Cu$_{7}$O$_{20+}$ 
& Copper-oxide superconductors    \\
\cline{1-2}
163 & (Sn$_{1.0}$ Pb$_{0.5}$ In$_{0.5}$)Ba$_4$Tm$_4$Cu$_{6}$O$_{18+}$ & \\
\cline{1-2}
& & \\
\cline{1-2}
125  & Tl$_2$Ba$_2$Ca$_2$Cu$_3$O$_{10}$ & \\
\cline{1-2}
108 & Tl$_2$Ba$_2$CaCu$_2$O$_8$ & \\
\cline{1-2}
80 & Tl$_2$Ba$_2$CuO$_6$ & \\
\cline{1-2}
& & \\
\cline{1-2}
110 & Bi$_2$ Sr$_2$ Ca$_2$ Cu$_3$ O$_{10}$(Bi2223) & \\
\cline{1-2}
92 &  Bi$_2$Sr$_2$CaCu$_2$O$_2$ (Bi2212)& \\
\cline{1-2}
92 & YBa$_2$ Cu$_3$ O$_7$ (YBCO) & \\
\hline
57 & SmFeAs(O,F)=SmOFeAs & Iron-based superconductors \\
\cline{1-2}
44 & LaFeAs(O,F)=LaOFeAs & \\
\hline
18 & Nb$_3$ Sn & \\
\cline{1-2}
10 & NbTi & Metallic low-temp. superconductors \\
\cline{1-2}
4.2 & Hg & \\
\hline
\end{tabular}
\label{tableSc}

\end{document}